\documentclass[twocolumn,showpacs]{revtex4}

\usepackage{graphicx}

\begin{document}


\title{
Fermi Surface Effect on
Lorenz Number of Correlated Metal
}

\author{Takuya Okabe}
\affiliation{
Faculty of Engineering, Shizuoka University, 
3-5-1 Johoku, 
Hamamatsu 432-8561,Japan}

\date{\today}

\begin{abstract}
We 
investigate 
an effect that
an ideal Lorenz number $L_{\rm i}$ of correlated metal
shows peculiar Fermi surface dependence,
which is caused by the onset of 
a particular channel of Umklapp scattering.
We evaluate $L_{\rm i}$ for some simple models and 
transition metals, and note 
that $L_{\rm i}$ for Na$_x$CoO$_2$ decreases sensitively 
as $x$ approaches an Umklapp threshold around $x_c \simeq 0.6$.
\end{abstract}

\maketitle

\section{Introduction}

The discovery of large thermopower in Na$_x$CoO$_2$\cite{tsu97}
has prompted quest for related good thermoelectric (TE) materials
and encouraged experimental and theoretical
research on TE properties of
strongly correlated electron systems.
In practical TE application, 
the important material parameter is
the dimensionless figure of merit $ZT =S^2/(\kappa/\sigma T)$,
where $S$ is the Seebeck coefficient,
$\kappa$ the thermal conductivity,
$\sigma$ the electrical conductivity,
and $T$ the absolute temperature.
Hence it is equally as important to 
enhance $S$ as it is to decrease
the Lorenz number $L$ defined by $L =\kappa/\sigma T$\cite{mahan98}.
The latter however is usually hampered by 
the Wiedemann-Franz (WF) law, according to which $L$ 
should be a universal constant,
e.g., $L=\frac{\pi^2}{3}(k/e)^2$ for impure metals.
In fact, normal metals, which generally have low $S$,
will not be good TE devices with $ZT>1$, 
unless the WF law is overcome.
It would thus be interesting to investigate 
a possible material dependence of $L$
to see if it is a controllable variable in principle.

We discuss the {\it ideal} value $L_{\rm i}$ 
(which is simply denoted as $L$ below)
of a correlated electron system on a
rigid lattice without impurity in the low temperature limit
for the purpose of elucidating 
its material specific dependence.
Indeed, for pure transition metals, the Lorenz numbers
have been observed to vary from metal to
metal\cite{schriempf68,wt67,apr68,btc70,wgb71}. 
Theoretically, 
Herring had derived a universal constant\cite{herring}, while 
the problem had also been addressed specifically by treating $s$-$d$
hybridization as perturbation, namely, by a two band model
of conductive $s$ electrons scattering off localized $d$ 
states\cite{rice68,ssm69}.
Herring's argument
based on the Fermi liquid theory
should be the proper approach to the problem
at low temperatures.
%
We derive a dimensionless factor (Eq.~(\ref{calL})) 
so as to modify his constant result, 
with which we investigate 
the Fermi surface (FS) dependence of $L$ 
for some typical cases concretely.
After showing nontrivial results obtained for simple models,
we give results of numerical evaluation of
the Lorenz numbers of some transition metals.
Lastly, 
we find it interesting to investigate
a model of the sodium 
cobalt oxide Na$_x$CoO$_2$,
because a nontrivial $x$ dependence of $L$
is expected theoretically 
owing to its simple cylindrical FS\cite{singh00}.



%
%
%
%

%
%
%

\section{Fermi Liquid theory}
\label{sec:FLT}

We begin with the linearized transport equation of 
a Fermi liquid
under the temperature gradient $\nabla T$ and
the electric field ${\bf E}$,
\begin{equation}
 \frac{\partial n_p}{\partial \varepsilon_p} 
\left(
-\frac{\varepsilon_p -\mu}{T}
\nabla T
\cdot {\bf v}_p
+ e{\bf E}\cdot {\bf v}_p 
\right)
=I[\varphi],
\label{lbeq}
\end{equation}
\begin{eqnarray}
TI[\varphi] &= &
\sum_{p',k} W^{pp'}_{k}
n_p n_{p'} (1-n_{p-k})(1-n_{p'+k}) (\varphi_p+\varphi_{p'}
\nonumber\\&&
- \varphi_{p-k}-\varphi_{p'+k}
)
\delta(\varepsilon_p+\varepsilon_{p'}-\varepsilon_{p-k}-\varepsilon_{p'+k}),
\nonumber
\end{eqnarray}
where
$n_p =n(\varepsilon_p) =1/({\rm e}^{(\varepsilon_p-\mu)/T}+1)$ is 
the Fermi distribution function,
and $W^{pp'}_{k}$ represents the transition probability
of quasiparticle scattering; $p, p' \rightarrow p-k, p'+k$.
In terms of the solution $\varphi_p$ of Eq.~(\ref{lbeq}),
the electric and heat currents 
carried by quasiparticles
are respectively given by 
\[
{\bf  J}=2 e
 \sum_p {\bf v}_p 
\frac{\partial n}{\partial \varepsilon_{p}} \varphi_p,\quad
 {\bf Q}=2
\sum_p {\bf v}_p (\varepsilon_p -\mu)
\frac{\partial n}{\partial \varepsilon_{p}} \varphi_p,
\]
where $\varepsilon_p$ and ${\bf v}_p$ are
energy and velocity of quasiparticles,
respectively.
In comparison with the phenomenological formulae
\(
{\bf J}=\sigma {\bf E}-\sigma S\nabla T
\) and 
\({\bf Q}=T\sigma S {\bf E}-\kappa_0 \nabla T,
\)
we obtain 
the transport coefficients $\sigma$, $S$, and $\kappa_0$,
the thermal conductivity at zero electric field.
The thermal conductivity at zero current, $\kappa$, is given by 
\(
 \kappa =\kappa_0 -T\sigma S^2
\)\cite{mahan98}.
Formally, the above results 
expressed in terms of the {renormalized} quantities
$\varepsilon_p$ and ${\bf v}_p$
bear resemblance to those of a weakly interacting Fermi gas.
Nevertheless, many body effects 
are included not only in the renormalization for the individual quasiparticle,
but also in 
the field induced shift of the quasiparticle distribution $\varphi_p$, 
which implicitly includes a collective Fermi liquid effect
depending on Landau parameters\cite{PN66,cond2}.
We do not write down the explicit expressions for them
as they are irrelevant for our purposes in what follows.


For definiteness, let us assume that
the currents ${\bf J}$ and ${\bf Q}$ flow in the $x$-direction.
Then, for the two functions  $l_1$ and $l_2$ defined by
\(
 \varphi_p =
- 
eE_x l_1 +{\partial_x T}  l_2
\), we obtain the equations
\begin{equation}
-\frac{\partial n_p}{\partial \varepsilon_p}
 {v}_{p x} 
=
I[l_1],
\label{eql1}
\end{equation}
and
\begin{equation}
-\frac{\partial n_p}{\partial \varepsilon_p}
t_p
 {v}_{p x} 
=
I[l_2],
\label{eql2}
\end{equation}
where $t_p=(\varepsilon_p-\mu)/T$.
For an isotropic system,
the collision integral $I$ has been evaluated analytically 
by an elaborate approximation\cite{ak59}.
As we cannot assume 
the predominance of normal scattering processes
in general cases of anisotropic Fermi liquids of our concern,
we have to make approximations in another way.
To simplify the multiple momentum sum in $I$,
we replace
\(
(1-n(\varepsilon_{p'+k}))
\delta(\varepsilon_p+\varepsilon_{p'}-\varepsilon_{p-k}-\varepsilon_{p'+k}) 
\)
with
\(
(1-n(\varepsilon_p+\varepsilon_{p'}-\varepsilon_{p-k})
)
\delta(\varepsilon_{p'+k}-\mu),
\)
for significant contributions to the collision term 
should come from a thermal neighborhood of the FS in any case.
Moreover, 
as in the isotropic case,
to describe the momentum dependence of the solutions $l_{i}$,
we 
decouple the crystal momentum variable ${\bf p}$ into 
the radial (energy) direction
$t_p=(\varepsilon_p-\mu)/T$
and the perpendicular component $\bar{\Omega}_p$,
and set
\(
 l_{i,p}=M_i(\bar{\Omega}_p)N_i(t_p)
\)
($i=1,2$).
Then the momentum sum is written as
\(
 \sum_{p} =
T\int \rho  {\rm d} t
 \int {\rm d}\bar{\Omega}_{p},
\)
where $\rho$ represents the density of quasiparticle states (DOS)
under the normalization condition $\int {\rm d}\bar{\Omega}_{p}=1$.
By integrating over the energy variables,
we obtain equations for $M_i(\bar{\Omega}_p)$.


First we note that 
we can reproduce the same conductivity formula
as derived and discussed previously\cite{YY86,cond2}.
By setting $N_1(t_p)$ as a constant,
the energy integrals which appear in 
the four terms in $I[l_1]$ of Eq.~(\ref{eql1})
can be calculated analytically. As a result, 
we obtain
\begin{widetext}
\begin{eqnarray}
 v_{p x}
&=&
\frac{\pi^2 }{2}\rho^2 T^2
\int {{\rm d}\bar{\Omega}_{p'}}
\int {{\rm d}\bar{\Omega}_{p''}}
W^{pp'}_{p-p''}
({M}_{1,p}+{M}_{1,p'}- {M}_{1,p''}-{M}_{1,p+p'-p''}
)
\delta(\varepsilon_{p+p'-p''}-\mu)
\nonumber\\
&=&
\frac{\pi^2 T^2}{2} 
\sum_{p',p''}
W^{pp'}_{p-p''}
({M}_{1,p}+{M}_{1,p'}- {M}_{1,p''}-{M}_{1,p+p'-p''}
)
\rho_{p'}\rho_{p''}\rho_{p+p'-p''},
\label{eqforM1}
\end{eqnarray}
\end{widetext}
and 
\begin{equation}
 \sigma = 
  {2e^2}
\rho \int {v}_{p x}  
M_1(\bar{\Omega}_p){{\rm d}\bar{\Omega}_p}.
\end{equation}
In Eq.~(\ref{eqforM1}), 
we used $\rho_p=\delta(\varepsilon_{p}-\mu)$.
Assuming the angular dependence
${M}_{1,p}\propto v_{p x}$, 
we obtain the resistivity coefficient 
\begin{equation}
 A = \frac{
\pi^2 \sum_{p,p',p''}
W^{pp'}_{p-p''}
\rho_{p}\rho_{p'}\rho_{p''}\rho_{p+p'-p''}
 v_{p_{x}}
({v}_{p{x}}+{v}_{p'{x}}- {v}_{p''{x}}-{v}_{p+p'-p'' x}
)
}{  
4 e^2\left(
\sum_p \rho_p  {v}_{p x} ^2 
\right)^2
}
\label{A=fracpi^2sum...}
\end{equation}
for the electrical resistivity $\sigma^{-1}=  A  T^2$.

On the other hand, 
we have to make a further approximation for $\kappa_0$.
%
By taking the $t$-derivative at $t_p=0$ of Eq.~(\ref{eql2}),
while adopting $N_2(t)\propto t/(t^2+\pi^2)$\cite{herring},
we finally obtain similar equations, namely,
\begin{eqnarray}
 v_{p x}
&=&
\rho^2 T^2
\int {{\rm d}\bar{\Omega}_{p'}}
\int {{\rm d}\bar{\Omega}_{p''}}
W^{pp'}_{p-p''}
({M}_{2,p} -c ({M}_{2,p'}
\nonumber\\&&
+ {M}_{2,p''}+{M}_{2,p+p'-p''}
)
)
\delta(\varepsilon_{p+p'-p''}-\mu),
\label{eqforM2}
\end{eqnarray}
where
\[
 c= \int_{0}^\infty {\rm d}t
\left(
{t} \coth \frac{t}{2} -2
\right)
\frac{2t/(t^2+\pi^2)}{\sinh {t}}
\simeq 0.162,
\]
and 
\(
 \kappa_0 = 
2 \pi^2  \rho T 
 \int_{-\infty}^\infty \frac{\partial n}{\partial t} \frac{t^2 {\rm d}t}{t^2+\pi^2}
 \int {v}_{p x}  
M_2(\bar{\Omega}_p){{\rm d}\bar{\Omega}_p},
\)
from which 
we obtain the thermal resistivity
$\kappa_0^{-1}= B T$.
It is remarked that
the difference in the integrand kernels of
Eqs.~(\ref{eqforM1}) and (\ref{eqforM2})
stems from the fact that 
$N_2(t)$ for $\kappa_0$ 
is an odd function, while $N_1(t)$ for $\sigma$ is even.
To evaluate the Lorenz number
$L= {\kappa}/{\sigma T} $
concretely, 
assuming $M_{2,p}\propto v_{p x}$ as above, 
we obtain 
\begin{equation}
L
< 
L_0\equiv  \frac{\kappa_0}{\sigma T} = 
\frac{\pi^2}{12}(12-\pi^2)
{\cal L} 
\left(
\frac{k}{e}\right)^2,
%
\label{L0}
\end{equation}
where the Boltzmann constant $k$ is written explicitly\footnote{
In the same manner, we get
\(
 \sigma S =
\frac{2\pi^2 e}{3}
T \int
\frac{{\rm d} (\rho v_{p x} )}{{\rm d}\varepsilon}
 M_1 {\rm d}\bar{\Omega}_p.
\)
For the parabolic band, we obtain
$S =\frac{2\pi^2 }{3e} \frac{\rho'}{\rho}T
=\frac{\pi^2}{3e} \frac{T}{\varepsilon_F} 
$,
the same 
result 
as that for free electrons.}.
The constant prefactor 
in Eq.~(\ref{L0})
is separated as it comes from the energy integral,
and corresponds to the result first derived by
Herring\cite{herring},
though our result differs by $\pi^2/12$.
In addition,
the dimensionless 
factor ${\cal L}$ in Eq.~(\ref{L0})
originates from 
the directional dependence of scattering,
and is given explicitly by
\begin{widetext}
\begin{eqnarray}
{\cal L} 
= \frac{A}{B}
&=&
 \frac{
\displaystyle
\int {{\rm d}\bar{\Omega}_1}
\int {{\rm d}\bar{\Omega}_2}
\int {{\rm d}\bar{\Omega}_3}
W^{p_1 p_2 }_{p_1-p_3}
 v_{p_{1}x}
({v}_{p_{1}x}+{v}_{p_{2}x}- {v}_{p_{3}x}-{v}_{p_1+p_2-p_3,x}
)
\delta(\varepsilon_{p_1+p_2-p_3}-\mu) 
}{
\displaystyle
\int {{\rm d}\bar{\Omega}_1}
\int {{\rm d}\bar{\Omega}_2}
\int {{\rm d}\bar{\Omega}_3}
W^{p_1 p_2 }_{p_1-p_3}
 v_{p_{1x}}
({v}_{p_{1}x}-
c\left(
{v}_{p_{2}x}+ {v}_{p_{3}x}+{v}_{p_1+p_2-p_3,x}\right)
)
\delta(\varepsilon_{p_1+p_2-p_3}-\mu) 
}
\label{calL0}
\\
&=&
 \frac{
\displaystyle
\sum_{1,2,3}
W^{p_1 p_2 }_{p_1-p_3}
\rho_{p_1}\rho_{p_2}\rho_{p_3}
\rho_{p_1+p_2-p_3}
 v_{p_{1}x}
\left(
{v}_{p_{1}x}+{v}_{p_{2}x}- {v}_{p_{3}x}-{v}_{p_1+p_2-p_3,x}
\right)
}{
\displaystyle
\sum_{1,2,3}
W^{p_1 p_2 }_{p_1-p_3}
\rho_{p_1}\rho_{p_2}\rho_{p_3}
\rho_{p_1+p_2-p_3}
 v_{p_{1}x}
({v}_{p_{1}x}
-c\left(
{v}_{p_{2}x}+ {v}_{p_{3}x}+{v}_{p_1+p_2-p_3,x}
\right)
)
}.
\label{calL}
\end{eqnarray}
\end{widetext}

Note that one would find
a trivial result ${\cal L}=1$, 
a constant Lorenz ratio, 
if one neglects both of 
the three vertex correction terms
following ${v}_{p_{1}x}$ in the parentheses 
of the denominator as well as the numerator of Eq.~(\ref{calL0}).
To have a nontrivial effect ${\cal L}\ne 1$, 
it is essential not to disregard detailed momentum structure of 
the relevant quasiparticle scatterings on the FS.
In fact, ${\cal L}$ 
quantifies the effect of FS geometry 
on the availability of phase space for the quasiparticle scatterings
to relax the transport currents.
Essentially, 
the above expression represents the fact that
thermal and electrical resistivities
are determined mainly by different types of scattering processes,
that is, 
normal processes are important for the thermal resistivity $B$,
while the electrical $A$ is caused by Umklapp processes.
In fact, for the numerator, or
the resistivity coefficient $A$,
to take a finite value,
there must exist at least
a set of four momenta ${\bf p}_i$ ($i=1,2,3,4$) 
on the FS
satisfying the Umklapp condition
${\bf p}_1+ {\bf p}_2 ={\bf p}_3+{\bf p}_4 +{\bf G}$,
where ${\bf G} (\ne 0)$ is a reciprocal lattice vector.
This is not met if the FS is too small.
Indeed, there is a limit Fermi radius estimated by
the relation $|{\bf p}_i| =|{\bf G}|/4$ which holds
at the threshold
where all the vectors are parallel or antiparallel.
In the vicinity of the threshold, 
one should expect a strong FS dependence of ${\cal L}$.
This is a matter of our concern in the following.

%
%
%

\section{results}
\label{sec:RES}

\subsection{Simple Model}

The factor ${\cal L}$ may be sensitive to
the momentum dependence of the scattering probability
\(
W^{pp'}_k=
\frac{2\pi}{\hbar} 
\left(
|A_{\uparrow\downarrow,k}^{pp'}|^2 
+\frac{1}{2}|A_{\uparrow\uparrow,k}^{pp'}|^2
\right),
\)
where 
$A_{\uparrow\uparrow,k}^{pp'}$ and
$A_{\uparrow\downarrow,k}^{pp'}$
are the scattering amplitudes for quasiparticles with parallel and
antiparallel spins.
In particular, the momentum dependence can 
give rise to a conspicuous effect
in the vicinity of quantum critical points
where there are quantum fluctuations localized in $k$ space.
To take this into account,
we may assume
\begin{equation}
A_{\uparrow\uparrow,k}^{pp'}\propto
A_{\uparrow\downarrow,k}^{pp'}
\propto
A_k^{pp'} = 
\frac{1}{{1+ \xi^2 \gamma({{\bf k}-{\bf Q}})}}
\label{modelAmp}
\end{equation}
in order to represent 
the effect of fluctuations peaked around ${\bf k}={\bf Q}$.
In addition to the correlation length $\xi$,
one may adopt
\(
 \gamma({\bf k}) \propto \sum_{{\bf d}}
\left(
1- {\rm e}^{{\rm i}{\bf k}\cdot {\bf d}}
\right)
\), where the sum is taken over the nearest neighboring lattice vectors
${\bf d}$,
as it gives
a simple lattice periodic function to give $\gamma({\bf k}) \simeq k^2$ 
in the long wavelength limit $k\rightarrow 0$.

To illustrate how ${\cal L}$ varies,
let us first investigate a simple two-dimensional (2D) model.
We should make a special remark, however.
In a strictly 2D system, one will find that
normal forward scatterings make
the denominator $B$ of Eq.~(\ref{calL0})
logarithmically divergent, as so for
the inverse lifetime of quasiparticle
which corresponds to the denominator of Eq.~(\ref{calL0}) with
$c=0$\cite{hsw71}.
On the other side, 
the numerator $A$ still remains finite
as noted by Fujimoto $et$ $al.$\cite{fky91}.
In effect, the divergence is suppressed by
a small decay rate $\Gamma$ assigned to the quasiparticle states,
$\rho_p =\delta (\varepsilon_p -\mu)$.
In a real system, such a cutoff must be provided by 
an inevitable effect of three dimensionality of the system
or by a finite density of impurities.
As the dependence on $\Gamma$ is logarithmic and weak numerically, 
here we present a typical behaviour 
assuming $\Gamma$ as a given constant for simplicity.

\begin{figure}
\includegraphics{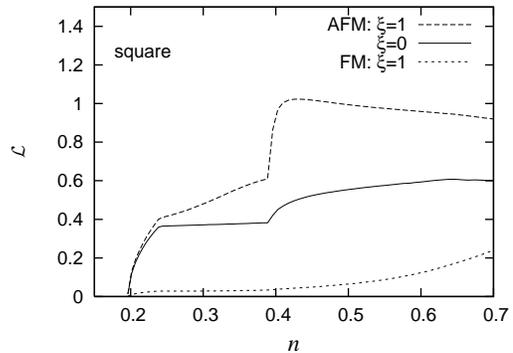}
\caption{\label{fig:square-qcp} 
For a parabolic band on a square lattice,
the dimensionless Lorenz factor
${\cal L}$ in Eq.~(\ref{L0})
is shown
as a function of the filling $n$ per orbital.
For $\xi=0$ (solid line),
for $\xi=1$ and ${\bf Q}=(\pi,\pi)$ (dashed line, AFM),
and for $\xi=1$ and ${\bf Q
}=(0,0)$
(dotted line, FM).
In all cases, 
${\cal L}$ must
vanish 
at 
the threshold 
$n_{c,1}\simeq 0.2$.
}
\end{figure}
First we discuss the simplest case of 
a parabolic band on a square lattice.
It is easy to show that
the result is independent of the quasiparticle mass,
so that we may simply use $\varepsilon_p =p^2$.
We obtain Fig.~\ref{fig:square-qcp} for $\Gamma=0.02$,
which representatively shows ${\cal L}$ 
as a function of the electron number $n$ per orbital
for three types of $W^{pp'}_k$ according to Eq.~(\ref{modelAmp}).
As expected, characteristic anomalies are clearly observed.
The solid line for $\xi=0$ ($W^{pp'}_k=$const.)
typically shows the first onset of the Umklapp processes
involving the smallest reciprocal lattice vector ${\bf G}=(2\pi,0)$
at $n_{c,1}=\pi/16\simeq 0.20$,
as well as the second one at $n_{c,2}\simeq 0.39$ 
for ${\bf G}=(2\pi,2\pi)$.
It is noted that
the threshold fillings can be easily evaluated 
as they are geometrically determined by the given FS.
The dashed line 
indicates 
that the latter structure for $n>n_{c,2}$
is particularly emphasized by 
the commensurate antiferromagnetic (AFM) fluctuations
with ${\bf Q}=(\pi,\pi)$.
On the other hand,
the ferromagnetic (FM) fluctuations 
with ${\bf Q}=(0,0)$ generally suppress ${\cal L}$,
as they strengthen the relative weight of the normal processes 
contributing to the thermal resistivity $B$.

\begin{figure}
\includegraphics{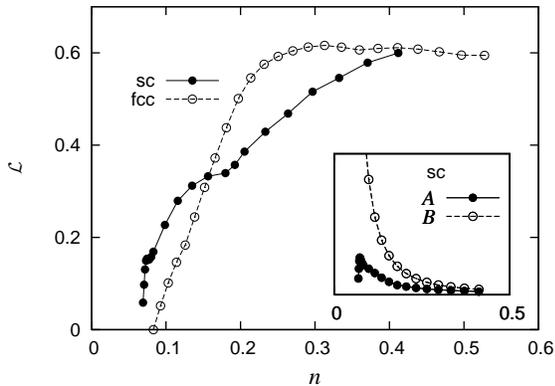}
\caption{\label{fig:scfcc}
${\cal L} (\equiv A/B)
$ 
of a parabolic band
on a 
sc
lattice
and a
fcc
lattice.
In the inset, 
the resistivity coefficients $A \propto \sigma^{-1}/T^2$ and $B$ 
for the sc lattice are shown as a function of $n$.
}
\end{figure}
Similarly, one may obtain
results for three dimensional systems,
in which no cutoff is required.
In Fig.~\ref{fig:scfcc}, 
we observe
the lattice-structure dependence of ${\cal L}$ 
for the parabolic band
in a simple cubic (sc) lattice and
a face centered cubic (fcc) lattice
with $W_k^{pp'}=$const. 
For the sc lattice the first threshold lies
at $n_{c,1}\simeq 0.065$ due to ${\bf G}=(2\pi,0,0)$,
while it is
at $n_{c,1}\simeq 0.085$ with ${\bf G}=(2\pi,2\pi,-2\pi)$
for the fcc lattice.
For the latter, the secondary kinks 
expected at $n_{c,2}\simeq 0.13$ and $n_{c,3}\simeq 0.37$ 
corresponding to ${\bf G}=(4\pi,4\pi,0)$ and $(4\pi,0,0)$, respectively,
are not so conspicuous as that found clearly at $n_{c,2}\simeq 0.19$ 
due to ${\bf G}=(2\pi,2\pi,0)$ for the sc lattice.
In the inset of Fig.~\ref{fig:scfcc},
the resistivity coefficients 
$A$ and $B$ 
are shown respectively for the sc lattice.
We remark that
the presence of the threshold $n_{c,1}$ may be more easily 
anticipated from a
relatively gradual $n$-dependence of ${\cal L}$
than from the electrical resistivity coefficient $A$,
which drops abruptly at $n_{c,1}$.
%
The results exemplify that 
the ideal Lorenz ratio is not a constant number
but shows the lattice structure dependence interestingly.


\subsection{Transition metals}

\begin{table}
\caption{\label{tab:table1}
Calculated values for the ideal Lorenz ratio
are compared with experiments.
}
\begin{ruledtabular}
\begin{tabular}{lccc}
& ${\cal L}$ &$L_0$ {($10^{-8}$V$^2$/K$^2$)} & $L_{\rm exp}$ {($10^{-8}$V$^2$/K$^2$)}\\
\hline
Pd ($\xi=0$)  & 1.0 & 1.3 &\\
Pd ($\xi=5$\AA) & 0.60 & 0.78 & 1.1 \cite{schriempf68} \\
Ni & 1.1 & 1.4 & 1.0 \cite{wt67} \\
Pt & 1.0 & 1.3&0.1 \cite{apr68} \\ 
Fe & 0.87& 1.1& 1.1 \cite{btc70} \\
W & 0.61& 0.79& 0.2-0.4 \cite{wgb71} 
\end{tabular}
\end{ruledtabular}
\end{table}
In principle, 
we can evaluate ${\cal L}$ for real materials
though it would generally require a hard task numerically.
Let us evaluate them for transition metals
 in the same manner as described elsewhere\cite{okabe07},
i.e., 
from the result of a first principle band calculation,
we pick up a main band with the largest DOS,
for which we apply Eq.~(\ref{calL}).
Table~\ref{tab:table1} shows
calculated values along with experimental 
results\cite{schriempf68,wt67,apr68,btc70,wgb71}. 
For these to be evaluated, we regarded $W_k^{pp'}$ as a constant,
except Pd for which presented also is the result with
a paramagnon ferromagnetic correlation effect with
the correlation length $\xi=5$\AA\cite{hhl68},
which is taken into account as above
in Eq.~(\ref{modelAmp}).
Considering the approximations made to derive Eq.~(\ref{calL}),
for the scattering amplitudes and so on,
we conclude that we could explain a small $L_{\rm exp}$ of
tungsten, among others.
This must be primarily due to a peculiar FS\cite{loucks65}.
For platinum, however, the observed value\cite{apr68} 
is inexplicable by our single band result.

%
%
%

\begin{figure}
\includegraphics{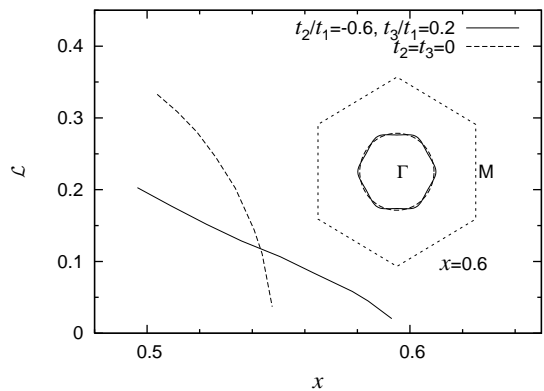}
\caption{\label{fig:triangle} 
For a single band 
tight-binding model of Na$_x$CoO$_2$,
${\cal L}$ as a function of $x$
as well as the Fermi surfaces at $x=0.6$
are shown to indicate 
the strong effect caused by 
a slight deformation due to the hopping integrals $t_2$ and $t_3$.
}
\end{figure}

\subsection{Na$_x$CoO$_2$}

Lastly, we discuss a tight-binding model of Na$_x$CoO$_2$.
Though this material has attracted much interest by its large
thermopower, 
it is of particular interest for us because of its simple cylindrical FS.
As a nearly 2D system on a triangular lattice, 
it has an almost circular hole surface centered at the $\Gamma$ point
with a slight hexagonal distortion 
bulging into the sides of the hexagonal Brillouin
zone\cite{singh00,qhw06,gbb07}.
Therefore, it provides us with an unique case
to realize the nontrivial effect as discussed above.
Indeed 
the FS suggests that the Umklapp threshold must exist
around $x_c\sim 0.6$, and 
one should expect that even a slight change in the shape of the FS 
around the threshold could 
have a striking effect on the doping dependence of 
the genuine quasiparticle transport properties and 
the ideal Lorenz factor.
In fact, 
if we assume a undistorted parabolic band,
it is straightforward to show that we should obtain 
$x_{c}= 2n_{c,1}-1 \simeq 0.55$.
Nevertheless, 
experimentally,
the $T^2$ dependence of the electrical resistivity
has been observed up to $x\simeq 0.7$\cite{lth04,fww04},
around which, therefore, it must be necessary 
to take into account a three dimensional lattice distortion\cite{qhw06,gbb07}.
For definiteness and simplicity,
we restrict ourselves to the doping regime where
the three dimensional effect is irrelevant.
We do not consider 
incipient 
ferromagnetic in-plain correlations\cite{bct04},
as they would not modify the result qualitatively.
Below, 
we pay special attention to an expected nontrivial behavior that
the factor ${\cal L}$ should tend to vanish 
as we approach a threshold $x\alt x_c$.
We present results for a 2D tight-binding model 
on a triangular lattice, in which 
up to the third neighbor hopping integrals $t_i$ ($i=1,2,3$) are
considered as in Ref.~\cite{ka07}, where
Kuroki and Arita have discussed 
the thermopower and the electrical conductivity 
by a relaxation time approximation 
with a single time scale $\tau$.
Here the approximation must be abandoned from the outset, 
since otherwise 
one would only obtain a trivial result of a constant value
regardless of the FS, as mentioned above.


To see how ${\cal L}$ tends to vanish around $x\alt x_c$
qualitatively, 
we show Fig.~\ref{fig:triangle} in which
the $x$-dependence of ${\cal L}$ around $x\alt x_c$ 
is shown for $t_1>0$ and $\Gamma/t_1=0.02$,
along with the Fermi surface for $x=0.6$.
In comparison with the dashed line for $t_2=t_3=0$,
the solid line shows
the strong effect around $x\alt x_c$ 
due to a slight modification of the FS caused by $t_2$ and $t_3$.
In effect,
as the threshold value $x_c\sim 0.6$ itself depends sensitively on
portions of the FS closest to the Brillouin zone boundary,
the result 
cannot be regarded as a quantitative prediction.
%
%
%
%
%
%
Furthermore, unfortunately, 
it is not easy to compare this nontrivial prediction
directly with experimental results,
because the intrinsic electron term of the thermal resistivity
($\propto BT$)
for the perovskite oxides 
has been completely outweighed by
contributions due to phonons and impurities\cite{lth04,fww04}.
Those extrinsic terms have to be separated out properly 
 to verify the nontrivial filling dependence. 

\begin{figure}
\includegraphics{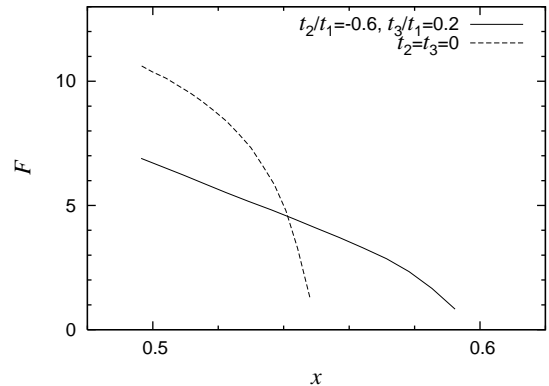}
\caption{\label{fig:triangleA} 
As in Fig.~\ref{fig:triangle}, 
shown are the filling $x$ dependences of the factor $F$,
proportional to the 
KW ratio $A/\gamma^2$.
}
\end{figure}
Lastly, 
as the vanishment of the factor ${\cal L}$  essentially
reflects that of the electrical resistivity,
it would be interesting also to calculate the $x$ dependence of 
the resistivity coefficient $A$,
or the Kadowaki Woods (KW) ratio $A/\gamma^2$.
To evaluate the resistivity coefficients themselves, however,
we need the absolute value of the scattering amplitude.
Assuming the strong coupling
\(
\rho^2 W^{pp'}_{p-p''} \simeq \pi
\)\cite{okabe07}, e.g., 
then we obtain 
\(
 A/\gamma^2 
= {9F }/{16 \pi e^2} 
\)
with 
\begin{eqnarray}
 F 
&=&
 \frac{
\displaystyle
\sum_{1,2,3}
\rho_{p_1}\rho_{p_2}\rho_{p_3}
\rho_{p_3-p_1-p_2}
 v_{p_{1x}}
({v}_{p_{1x}}+{v}_{p_{2x}}- {v}_{p_{3x}}-{v}_{p_1+p_2-p_3,x}
)
}{
\displaystyle
 \rho^4 
\left(
\sum_p
\rho_p {v}_{p x}^2 
\right)^2
}.
\label{factorF}
\end{eqnarray}
As in Fig.~\ref{fig:triangle} for ${\cal L}$, 
we show 
$F$ as a function of $x$ 
in Fig.~\ref{fig:triangleA}.
The ratio $A/\gamma^2$ also vanishes as $x$ approaches $x_c$,
as expected.
However, 
numerically, we find that it
should remain of the order of a common value, 
$A/\gamma^2\simeq 1\times 10^{-5} \mu\Omega$ cm(mol K/mJ)$^2$.
Li $et$ $al.$\cite{lth04} have observed 
a strongly enhanced deviation from this standard value
for $x=0.7$, which should be well beyond the threshold $x_c$.
It is difficult to regard
the numerical deviation simply as a FS effect.
Apart from taking into account a relevant three dimensional effect,
one would have to assume an enhanced scattering
\(
\rho^2 W^{pp'}_{p-p''} \gg 1, 
\)
which may be due to strong scatterings caused by a proximity to 
some sort of instability\cite{lth04}.

\section{Summary}
\label{sec:SUM}

In summary, 
on the basis of anisotropic Fermi liquid theory,
we investigated 
the ideal Lorenz ratio for correlated metals
by taking due care of the momentum dependence 
of transport relaxation processes due to 
mutual elastic scatterings between quasiparticles.
It was shown explicitly 
that the ideal Lorenz ratio of
a correlated electron system is not a constant, but 
may vary drastically
in circumstances and even be made vanishingly small,
not only by quantum fluctuations but by a filling control
around thresholds of Umklapp scattering channels.
Although it might not be easy to extract
the ideal Lorenz value in practice,
theoretically we pointed out and discussed that such a nontrivial 
effect should be expected in 
such a simple single-band system as Na$_x$CoO$_2$.

%



The numerical calculations were carried out 
on Altix4700 at Shizuoka University 
Information Processing Center.

\bibliography{okabe}

\end{document}